\documentclass[journal, onecolumn]{IEEEtran}
\IEEEoverridecommandlockouts

\usepackage{cite}
\usepackage{amsmath,amssymb,amsfonts}
\usepackage{enumitem}
\usepackage{algorithmic}
\usepackage{graphicx}
\usepackage{textcomp}
\usepackage{tikz}
\usepackage{epstopdf}
\usepackage{tabularx}
\newcolumntype{M}[1]{>{\centering\arraybackslash}m{#1}}
\usepackage[utf8]{inputenc}

\usepackage{arydshln,leftidx,mathtools}
\usepackage{bm}
\usetikzlibrary{spy}
\usepackage{xcolor}
\def\BibTeX{{\rm B\kern-.05em{\sc i\kern-.025em b}\kern-.08em
    T\kern-.1667em\lower.7ex\hbox{E}\kern-.125emX}}
    
\graphicspath{{Images/}}
\begin{document}

\title{2D PET Image Reconstruction Using Robust $L_1$ Estimation of the Gaussian Mixture Model \\
\thanks{This work has been fully supported by Croatian Science Foundation under the project IP-2014-09-2625.}
}

\author{\IEEEauthorblockN{Azra Tafro, Damir Seršić and Ana Sović Kržić}
\IEEEauthorblockA{\textit{Department of Electronic Systems and Information Processing} \\
\textit{University of Zagreb, Faculty of Electrical Engineering and Computing}\\
Zagreb, Croatia \\
azra.tafro@fer.hr, damir.sersic@fer.hr, ana.sovic.krzic@fer.hr}
}

\maketitle

\begin{abstract}
An image or volume of interest in positron emission tomography (PET) is reconstructed from pairs of gamma rays emitted from a radioactive substance. Many image reconstruction methods are based on estimation of pixels or voxels on some predefined grid. Such an approach is usually associated with limited resolution of the reconstruction, high computational complexity due to slow convergence and noisy results. 
This paper explores reconstruction of PET images using the underlying assumption that the originals of interest can be modeled using Gaussian mixture models. 
A robust segmentation method based on statistical properties of the model is presented, with an iterative algorithm resembling the expectation-maximization algorithm. Use of parametric models for image description instead of pixels circumvent some of the mentioned limitations.
\end{abstract}


\section{Introduction}
Positron emission tomography (PET) scanners detect the annihilation photon pairs arising from the positron emissions.  Image reconstruction implies generating a two- or three-dimensional image of a radiotracer's concentration to estimate physiologic parameters for volumes of interest in vivo. In three-dimensional scanners, we can consider the tube (parallelepiped) joining any two detector elements as a volume of response (VOR). In the absence of physical effects 
and noise, the total number of coincidence events detected will be proportional to the total amount of tracer contained in the volume of response. In the two-dimensional case, we consider only lines of response (LORs) joining two detector elements, lying within a specified imaging plane. The data are recorded as event histograms (sinograms or projected data) or as a list of recorded photon-pair events (list-mode data). For a general overview of standard PET image reconstruction methodology, see e.g. \cite{Tong2010ImageRF, alessio2006pet} or  \cite{Reader2007}.

Modern image reconstruction methods are mostly based on maximum-likelihood expectation-maximization (MLEM) iterations. Maximum likelihood is used as the optimization criterion, combined with the expectation-maximization algorithm for finding its solution. To overcome the computational complexity and slow convergence of the MLEM, the ordered subsets expectation-maximization (OSEM) algorithm has been recently introduced. Since MLEM or OSEM estimation of pixels or voxels is usually noisy \cite{Tong2010ImageRF}, use of post-filtering methods is necessary. On the one hand, image reconstruction from its projections is a mature research field with well known methods and proven results. On the other hand, known limitations made a challenge for a different approach presented in this paper.

In image segmentation, a number of algorithms based on model-based techniques utilizing prior knowledge have been proposed to model uncertainty (see e.g. \cite{Zhang2001}, \cite{Zhang1994}). The Gaussian mixture model (GMM) is a well-known and widely used model in a variety of segmentation and classification problems (\cite{Friedman1997},\cite{Ralasic2018}).

In this paper, we propose a new robust algorithm for efficient reconstruction of an object from a PET image. Our method utilizes a novel $L_1$ minimizing algorithm to estimate the parameters of a Gaussian mixture model within framework similar to the standard expectation-maximization (EM) algorithm. Instead of voxels, we estimate parameters of the object's model. We focus on the two-dimensional case for clarity, but an extension to three dimensions would follow in a similar fashion. This paper is organized into six sections. Section~\ref{sec:2D} gives an overview of traditional methodology in 2D PET imaging. In Section ~\ref{sec:gmm} an introduction to Gaussian mixture models is presented. Section~\ref{sec:estimation} describes the estimation of GMM parameters from PET data. Section~\ref{sec:EM} shows the implementation of iterative estimates in the EM-like algorithm. Finally, experimental results with discussion conclude the paper.
\section{Two-Dimensional PET Imaging}\label{sec:2D}
\begin{figure}[t]
\begin{minipage}[t]{0.5\linewidth}
	{\includegraphics[width=\linewidth]{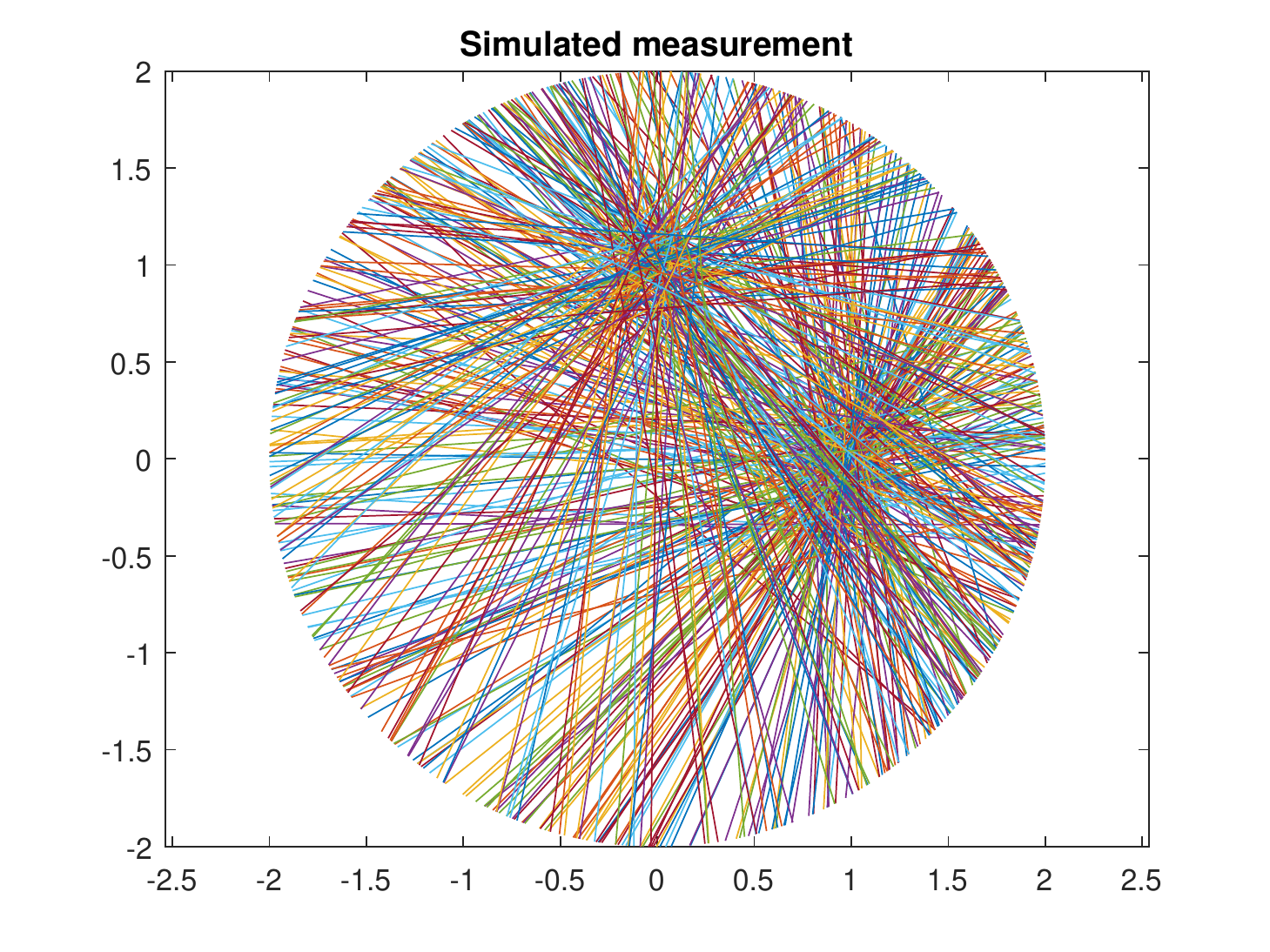}}
\end{minipage}%
    \hfill%
\begin{minipage}[t]{0.5\linewidth}
	{\includegraphics[width=\linewidth]{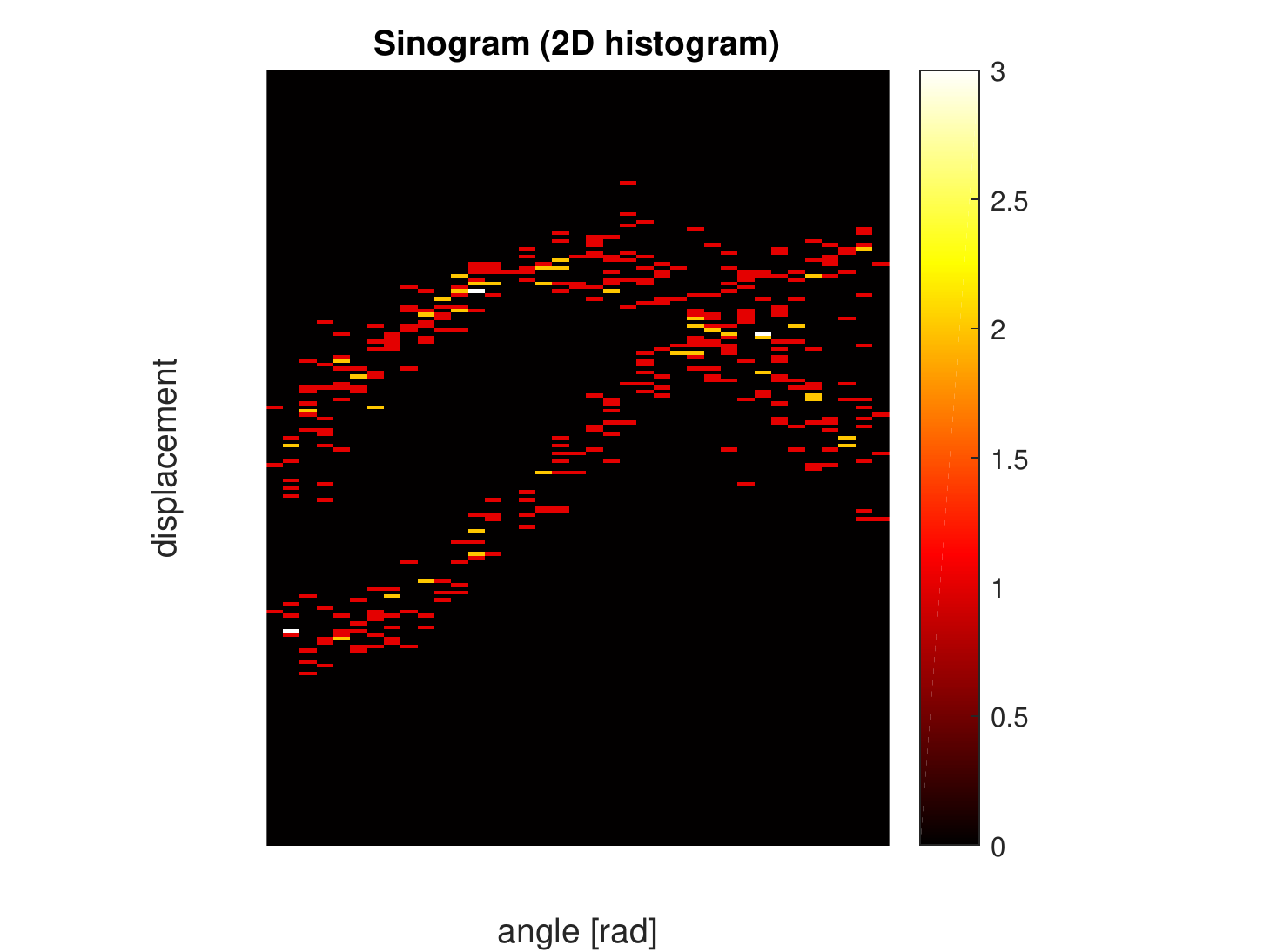}}
\end{minipage}
\caption{(\textbf{a}) Simulated measurements for $N=400$ LORs and $K=2$. (\textbf{b}) The corresponding sinogram.}
	\label{fig:measurement}
\end{figure}

In the two-dimensional case, the acquired data are collected along LORs through a two-dimensional object. Traditionally, data are organized into sets of projections, integrals along all LORs for a fixed direction $\phi$. The collection of all projections for $0\leq \phi<2\pi$ forms a two
dimensional function of the distance of the LOR from the origin, denoted by $s$, and angle $\phi$. The line-integral transform of $f(x,y) \to p(s,\phi)$ is called the X-ray transform \cite{Natterer1986}, which in 2D is the same as the Radon transform \cite{Helgason1999}. Since a fixed point traces a sinusoidal path in the projection space, the superposition of all sinusoids corresponding to each point of activity
in a general object is called a sinogram. This is illustrated in Fig.~\ref{fig:measurement}. 

A classical image reconstruction method is the filtered backprojection (FBP), which is based on the projection theorem of Fourier analysis. The algorithm reconstructs the image by calculating the inverse Fourier transform of the 2D Fourier transform of the backprojected image (see e.g. \cite{alessio2006pet}, \cite{Natterer1986}, \cite{OSullivan1993}). This technique does not include or depend on any assumptions about the data. Assumptions such as form, volume and dependence could be utilized to obtain more precise estimates. 

\section{Gaussian Mixture Models} \label{sec:gmm}
A Gaussian mixture model \cite{Reynolds2015} is a weighted sum of $K$ component Gaussian densities as given by the equation:
\begin{equation}
	p(\boldsymbol{x}\vert \tau_k, \boldsymbol{\mu}_k, \boldsymbol{\Sigma}_k )=\sum_{k=1}^{K}\tau_k~g(\boldsymbol{x}\vert \boldsymbol{\mu}_k, \boldsymbol{\Sigma}_k),
\end{equation}
where $\boldsymbol{x}$ is a $d$-dimensional observation, $\tau_i~(i=1,...,K)$ are the weights of each Gaussian component, and $g(\boldsymbol{x}\vert \boldsymbol{\mu}_k, \boldsymbol{\Sigma}_k)$ are the Gaussian component densities. We assume that a given observation~$\boldsymbol{x}$ is a realization from exactly one of the $K$ Gaussian mixture components, and each component density is a $d$-variate Gaussian function of the form:
\begin{equation} \label{Gauss_density}
	g(\boldsymbol{x}\vert \boldsymbol{\mu}_k, \boldsymbol{\Sigma}_k)=\frac{1}{\sqrt{(2\pi)^d\vert \boldsymbol{\Sigma}_k\vert}} \exp\Bigg(-\frac{1}{2}(\boldsymbol{x}-\boldsymbol{\mu}_k)^T \boldsymbol{\Sigma}_k^{-1}(\boldsymbol{x}-\boldsymbol{\mu}_k)\Bigg),
\end{equation}
with $\vert \boldsymbol{\Sigma}_k\vert$ denoting the determinant of $\boldsymbol{\Sigma}_k$. The set of probabilities $\{\tau_k\}$ such that $\sum_{k=1}^K\tau_k=1$ defines the probabilities that $\boldsymbol{x}$ belongs to the corresponding Gaussian component.
 The complete Gaussian mixture model is parameterized by the mean vectors $\{\boldsymbol{\mu}_k\}$, covariance matrices $\{\boldsymbol{\Sigma}_k\}$ and mixture weights $\{\tau_k\}$ for all Gaussian component densities. 

Traditionally, in image segmentation GMMs are used to model values at points of observation, e.g. activity concentration in voxels \cite{Layer2015} or pixel values \cite{Nguyen2013}. Those models do not take into account the spatial correlation between observations, which can be corrected by introducing Markov random fields \cite{Layer2015,Nguyen2013}. 

We propose applying the GMM directly to the spatial data, focusing only on the locations and not the values at those locations.

In this scenario, the $K$ Gaussian components represent potential points of origin for activity. The points $\boldsymbol{x}$ that are realizations from these components are latent (unobserved). Our observations are lines (LORs) through these points at random angles $\psi$, however using convenient properties of Gaussian distributions we will still be able to accurately estimate the parameters.

\section{Estimating Gaussian Parameters}\label{sec:estimation}
For clarity, in this section we focus only on one Gaussian component. Assuming we know a set of $N$ LORs whose points of origin come from a Gaussian distribution with parameters $(\boldsymbol{\mu}, \boldsymbol{\Sigma})$, we can estimate those parameters.
\subsection{Estimating $\bm{\mu}$ using minimal distance}
The mean vector $\boldsymbol{\mu}=[\mu_x ~ \mu_y]^T$ can be estimated in multiple ways. One method is to find the point in space whose total squared distance from all LORs is minimal. Clearly, the solution depends on our definition of distance. 

Each LOR in two dimensions is uniquely given by its slope $k=\tan\psi$ and an intercept $l$ (or by any two equivalent parameters) and we can write the LOR equations as
\begin{equation}\label{eq:LORs}
\bm{a}_i^T\bm{x}+l_i=0\,,\, \bm{a}_i=[\tan\psi_i~ {-}1]^T\, , \, i=1,\dots,N.
\end{equation}
In general, we can define the squared distance between $d$-dimensional vectors $\bm{v}_1$ and $\bm{v}_2$ as
\begin{equation}\label{eq:dist}
d^2(\bm{v}_1, \bm{v}_2)=(\bm{v}_1-\bm{v}_2)^T \bm{W} (\bm{v}_1-\bm{v}_2),
\end{equation}
for some $d\times d$ weight matrix $\bm{W} $. In particular, when $\bm{W} =\bm{I}$ the distance in \eqref{eq:dist} is the Euclidian distance, and for $\bm{W} =\bm{\Sigma}^{-1}$ we get the Mahalanobis distance. In our case $d=2$ and a planar line is given by one equation, but similar results would follow for higher dimensions. For a given $\bm{\mu}$, the point on the $i$-th line that is nearest to $\bm{\mu}$, i.e. the solution to 
\[\min d(\bm{x},\bm{\mu})\textrm{ s.t. }\bm{a}_i\bm{x}+l_i=0,\]
denoted by $\bm{x}_i^m$, can be found by solving \cite{Golub2012}
\begin{gather}\label{eq:xmin}
\begin{bmatrix}
2\bm{W} & \bm{a}_i  \\ 
\bm{a}_i^T & 0 
\end{bmatrix}\cdot\begin{bmatrix}
\bm{x}_i^m 	\\
\lambda	\\	
\end{bmatrix}
=
\begin{bmatrix}
2\bm{W}\bm{\mu}\\
-l_i
\end{bmatrix}.
\end{gather}
Now the point $\bm{\mu}$ that is nearest to all lines is the solution of 
\begin{equation}\label{eq:muproblem}
\min \sum_{i=1}^N(\bm{x}_i^m-\bm{ \mu})^T \bm{W}  (\bm{x}_i^m-\bm{\mu}),
\end{equation}
where $\bm{x}_i^m$ is given in \eqref{eq:xmin}. This can be shown to be (see Appendix~\ref{AppendixA}) the solution to:
\begin{equation}\label{eq:mu}
\left(\sum_{i=1}^N(2\bm{W} \bm{M}_i-\bm{N}_i)\right)\bm{\mu}^T=\sum_{i=1}^N l_i\bm{m}_i,
\end{equation}
where \begin{gather*}
\widetilde{\bm{W}}=
\begin{bmatrix}
\bm{W} & \bm{0} 	\\
\bm{0}	& 0\\	
\end{bmatrix}\in M_3\, , \, \bm{B}_i=\left(\begin{bmatrix}
2\bm{W}  	& \bm{a}_i  \\ 
\bm{a}_i^T & 0 
\end{bmatrix}^{-1}\right)^T\cdot \widetilde{\bm{W}}\in M_3,
\end{gather*}
and 
\begin{gather*}
\begin{bmatrix}
\bm{M}_i 	\\
\bm{m}_i	\\	
\end{bmatrix}
=\bm{B}_i\left(
\begin{bmatrix}
2\bm{W}  	& \bm{a}_i  \\ 
\bm{a}_i^T & 0 
\end{bmatrix}
\begin{bmatrix}
2\bm{W} 	\\
\bm{0}	\\	
\end{bmatrix}
- \begin{bmatrix}
1 & 0  	\\
0 & 1	\\	
0 & 0 \\
\end{bmatrix}\right),
\\
\begin{bmatrix}
\bm{N}_i 	\\
\bm{0}	\\	
\end{bmatrix}
=\bm{B}_i^T\begin{bmatrix}
2\bm{W} 	\\
\bm{0}	\\	
\end{bmatrix}-
\widetilde{\bm{W}}
\begin{bmatrix}
1 & 0  	\\
0 & 1	\\	
0 & 0 \\
\end{bmatrix},\, \bm{M}_i,\bm{N}_i\in M_2,\bm{m}_i\in M_{12}.
\end{gather*}
Note that the zeros and null-vectors added in expressions above appear for calculation purposes (see Appendix~\ref{AppendixA}) and do not change the original problem. 

It can also be shown that, for $\bm{\Sigma}$ known, the estimate obtained using the Mahalanobis distance is also the maximum likelihood estimate for $\bm{\mu}$.
\subsection{Estimating $\bm\Sigma$ using 1D projections}
In the two-dimensional setting, since $\bm{\Sigma}$ is a symmetric matrix, 
\[\bm{\Sigma}=\begin{bmatrix}
\Sigma_{11} & \Sigma_{12}\\
\Sigma_{12} & \Sigma_{22}
\end{bmatrix},\]
we need to estimate three parameters: $\Sigma_{11}$, $\Sigma_{12}$ and $\Sigma_{22}$.

Observe a single line of response, and recall that one of the parameters determining the LOR is the slope $\tan \psi$, where $\psi$ is the angle between the line and the $x$-axis. That means that by rotating the $xy$-coordinate system by $\psi-\frac{\pi}{2}$, the LOR would be parallel to the new $y$-axis. This is equivalent to the rotation of the Gaussian distribution by $\varphi=\frac{\pi}{2}-\psi$. This is illustrated in Fig.~\ref{fig:rotation}. Since a rotated Gaussian distribution is again Gaussian, in this new coordinate system the LOR is parallel to the $y$-axis, and the Gaussian distribution has parameters
\[(R\bm{\mu}, R\bm{\Sigma}R^T),\textrm{ where } R=\begin{bmatrix}
\cos\varphi & -\sin\varphi  	\\
\sin\varphi & \cos\varphi	
\end{bmatrix}.\] 
\begin{center}
\centering
\begin{figure}[t]
\centering
\includegraphics[scale=0.7]{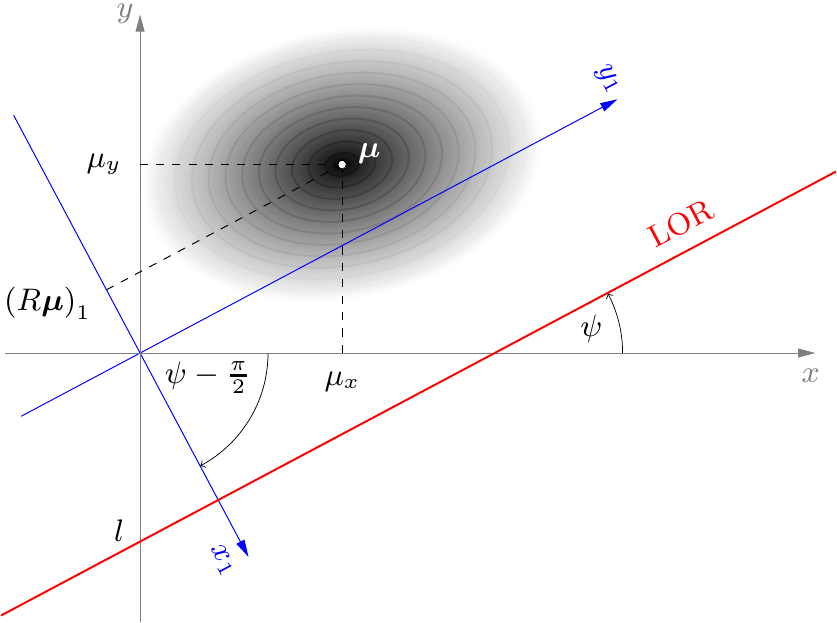}
\caption{Rotation of the coordinate system, $y$-axis parallel to the LOR.}\label{fig:rotation}
\end{figure}
\end{center}
Given the original LOR parameters, the coordinates of the point at which it intersects the new $x$-axis are $(-l\sin\varphi, 0)$. 
Since marginal distributions of a Gaussian are again Gaussian, a 1D projection onto the new $x$-axis is a Gaussian random variable with expectation
\[(R\bm{\mu})_1=\cos\varphi\mu_x-\sin\varphi\mu_y,\]
and variance
\begin{equation}\label{eq:sigmaest}( R\bm{\Sigma}R^T)_{11}=\cos^2\varphi \Sigma_{11}-2\cos\varphi\sin\varphi\Sigma_{12}+\sin^2\varphi\Sigma_{22},
\end{equation}
i.e. we obtain the one-dimensional mean and variance simply by omitting rows and columns from their multidimensional counterparts.

Therefore, each LOR gives us a one-dimensional projection whose squared (Euclidian) distance from the mean, $(\cos\varphi\mu_x-\sin\varphi\mu_y+l\sin\varphi)^2$, can be used to estimate the variance in \eqref{eq:sigmaest}. This gives us a system of equations:
\begin{equation}\label{eq:sigmamin}
\bm{A}\bm{s}=\bm{b},\textrm{ where }\bm{s}=\begin{bmatrix}
\Sigma_{11}\\
\Sigma_{12}\\
\Sigma_{22}
\end{bmatrix},
\end{equation}
\begin{gather*}
A=\begin{bmatrix}
\cos^2\varphi_1 & -2\sin\varphi_1\cos \varphi_1 & \sin^2\varphi_1 \\
\cos^2\varphi_2 & -2\sin\varphi_2\cos \varphi_2 & \sin^2\varphi_2 \\
\vdots & \vdots & \vdots\\
\cos^2\varphi_N & -2\sin\varphi_1\cos \varphi_N & \sin^2\varphi_N \\
\end{bmatrix}, 
\textrm{and }
\bm{b}=\begin{bmatrix}
(\cos\varphi_1\mu_x-\sin\varphi_1\mu_y+l_1\sin\varphi_1)^2\\
\vdots\\
(\cos\varphi_N\mu_x-\sin\varphi_N\mu_y+l_N\sin\varphi_N)^2
\end{bmatrix},
\end{gather*}
Since there are (dozens of) thousands of measurements, the problem in \eqref{eq:sigmamin} is overdetermined. It does not have an exact solution, but we can find the best approximation, i.e. the solution to 
\begin{equation}\label{eq:sigmamin2}
\min_{\bm{s}}\Vert \bm{A}\bm{s}-\bm{b}\Vert.
\end{equation}
Note that the solution will depend on the type of norm used in \eqref{eq:sigmamin2}. Following classical methodology, the ordinary least squares (OLS) method gives the solution that minimizes the $L_2$ norm. This solution can be found by solving the equivalent problem
\[\bm{A^T}\bm{A}\bm{s}=\bm{A^T}\cdot\bm{b}, \text{ i.e. }\bm{s}=(\bm{A^T}\bm{A})^{-1}\bm{A^T}\cdot\bm{b}.\]
As we will show in Sec.~\ref{sec:results}, OLS performs poorly when we introduce more than one component, so we will need to use alternative methods. It can be shown that in some cases $L_1$ minimization is preferred to the more traditional $L_2$ minimization because it is less sensitive, i.e. more resistant to gross and systematic errors \cite{Bektacs2010}. The main argument against $L_1$ minimisation would be computational complexity, which can be alleviated by using iterative methods. In this paper, the solution is obtained by using the $L_1$ minimization algorithm proposed in  \cite{Sovic2018}. We use it to solve a modified problem
\begin{equation}\label{eq:sigmaminL1}
\bm{A}\bm{s}=k\cdot\bm{b},
\end{equation}
where $k$ is a constant corrective scale factor.  
In a sufficiently large sample, one would have enough data points in each direction $\phi$ to obtain a one-dimensional variance estimate. In the absence of that,  we are able to obtain a robust estimator for the parameters of the covariance matrix following the median absolute deviation (MAD) estimator of deviation $\sigma$ \cite{Rousseeuw1993}, i.e.
\[k=(\Phi(0.75))^2\approx 1.4826^2,\]
where $\Phi$ denotes the distribution function of the standard normal random variable.

\section{Iterative EM-like Algorithm}\label{sec:EM}
The expectation-maximization (EM) algorithm, first explained in \cite{Dempster1977}, is an iterative method used to find maximum likelihood or maximum a posteriori (MAP) estimates of parameters in statistical models. The expectation (E) step creates a function for the expectation of the log-likelihood evaluated using the current estimate of parameters. The maximization (M) step then computes parameters maximizing the expected log-likelihood from the E step. Conversely, these estimates are then used in the next E step. In traditional applications of the EM algorithm to GMMs, the E step assigns each data point its \textit{membership probabilities} $\{\tau_k\}$, i.e. the probabilities that the point belongs to each of the mixture components. In the M step, parameters of each component are estimated using the points "belonging" to that component. As already mentioned, in the PET setting observations are lines, however we can replicate the iterative steps using estimates from Sec.~\ref{sec:estimation}. Alternatively, this could also be considered a Lloyd-like algorithm, where the difference from the conventional Lloyd's algorithm \cite{Lloyd1982} is that we allow different distance functions of the form \eqref{eq:dist}. The algorithm initializes parameters arbitrarily, and then alternates between the following steps:
\begin{enumerate}
\item Compute class membership probabilities. For each LOR, compute the squared distance from each component mean. We distinguish between a \textit{hard} classification where we assign the LOR to its nearest component, and a \textit{soft} classification where membership probability is inversely proportional to the squared distance.
\item Estimate component parameters. Either from a hard or soft classification, where each LOR participates with its proportional share, parameters of each component are estimated using methodology from Sec.~\ref{sec:estimation}.
\end{enumerate}
The $L_1$ minimization algorithm recursively reduces and increases dimensionality of the observed subspace and uses weighted median to efficiently find the global minimum, and has shown to overperform state-of-the-art competitive methods when there are relatively few parameters to be estimated from a very high number of equations. For details, see Appendix \ref{AppendixB} and  \cite{Sovic2018}.

Initial steps of the iterative algorithm use Euclidian distance in \eqref{eq:dist}. Since later iterations improve the estimates, the distance gradually transforms into the Mahalanobis distance, i.e.
\[\bm{W}=(1-\alpha)\bm{I}+\alpha\hat{\bm{\Sigma}}^{-1},\]
where $\alpha$ increases from $0$ to $1$. Therefore, in later iterations we obtain an MLE-like estimate of $\bm{\mu}$.

\section{Results and remarks}\label{sec:results}
To evaluate the methodology, we experimented in the two-dimensional setting with $K=1$ and $K=2$ components. 

First, for proof of concept we show that the method in Sec.~\ref{sec:estimation} provides good estimates with both $L_1$ and $L_2$ minimization, for several covariance matrices with varying corresponding correlation coefficients:
\[\bm{\Sigma}_{1}=0.05\bm{I},\, \bm{\Sigma}_{2}=\begin{bmatrix}
0.02 & -0.01\\
-0.01 & 0.05
\end{bmatrix}, \, \bm{\Sigma}_{3}=\begin{bmatrix}
0.01 & 0.02\\
0.02 & 0.05
\end{bmatrix}.\]
Since the 2D covariance matrix is of the form
\[\bm{\Sigma}=\begin{bmatrix}
\sigma_x^2 & \rho\sigma_x\sigma_y\\
\rho\sigma_x\sigma_y & \sigma_y^2
\end{bmatrix},\]
it is determined by three parameters -- $\sigma_x$, $\sigma_y$ and $\rho(=\rho_{xy})$. This corresponds to the three-dimensional vector \[\bm{s}=[\Sigma_{11} ~ \Sigma_{12} ~ \Sigma_{22}]^T=[\sigma_x^2 ~ \rho\sigma_x\sigma_y ~ \sigma_y^2]^T\] 
in \eqref{eq:sigmamin}. Instead of observing true and estimated $\bm{\Sigma}$ matrices we will calculate the error in estimation of $\bm{s}$ for each $\bm{s}$ that corresponds to matrices above:
\[\bm{s}_1=\begin{bmatrix}
0.05\\
0\\
0.05
\end{bmatrix},\,\bm{s}_2=\begin{bmatrix}
0.02\\
-0.01\\
0.05
\end{bmatrix},\,\bm{s}_3=\begin{bmatrix}
0.01\\
0.02\\
0.05
\end{bmatrix}.\] 
Note that the corresponding correlation coefficients can be calculated from these vectors, and they are $\rho_1=0$, $\rho_2\approx -0.3$, $\rho_3\approx 0.9$.

For each of these types we simulated a measurement from $N=1000$ and $N=10000$ points and calculated  the average estimate for $L_1$ and $L_2$ estimates separately. We repeated the experiment $1000$. Accuracy of an estimate can be assessed in many ways, for illustrative purposes we chose relative error i.e. 
$$\Vert \bm{e}\Vert= \frac{\Vert\hat{\bm{s}}-\bm{s}\Vert}{\Vert \bm{s}\Vert},$$ 
where $\Vert \cdot \Vert$ denotes the standard Euclidian norm in both expressions. Results are given in Table \ref{tab:results1}.
\begin{table}[h]
\renewcommand{\arraystretch}{1.2}
\centering
\caption{Mean estimation error, $K=1$}
\label{tab:results1}
\begin{tabular}{|c|c|c|c|}
\hline
$N=1000$ & $\bm{s}_1$ & $\bm{s}_2$ & $\bm{s}_3$ \\
\hline
$L_1$ method& 13.78\% & 11.88\% &  9.28\%\\ 
\hline
$L_2$ method& 8.27\% & 7.61\% &  7.6\%\\ 
\hline
\end{tabular}

\vspace{0.2cm}

\begin{tabular}{|c|c|c|c|}
\hline
$N=10000$ & $\bm{s}_1$ & $\bm{s}_2$ & $\bm{s}_3$ \\
\hline
$L_1$ method& 4.28\% & 3.66\% &  2.93\%\\ 
\hline
$L_2$ method& 2.61\% & 2.38\% &  2.37\%\\ 
\hline
\end{tabular}
\end{table}

Given that the variance of the traditional standard deviation estimator (from points) equals $\frac{\sigma^4}{n-1}$, estimations within $10\%$ from $N=1000$ LORs seem acceptable, which justifies the methodology described in Sec.~\ref{sec:estimation}. Significantly more accurate estimation from $N=10000$ LORs further confirms this. It is also notable that the accuracy of the estimate increases as $\vert \rho\vert|$ increases from $0$ to $1$. 


For $K=2$ components we repeated the experiment as described in Sec.~\ref{sec:EM} for various combinations of types of vector $\bm{s}$. We used hard classification, where we assign each line to at most one component. We experimented with various constraints, from simply assigning LORs to more likely components to assignations only when the probability of belonging is above a certain threshold.
 
\begin{figure}[ht]
\begin{minipage}[t]{0.5\linewidth}
	{\includegraphics[width=\linewidth]{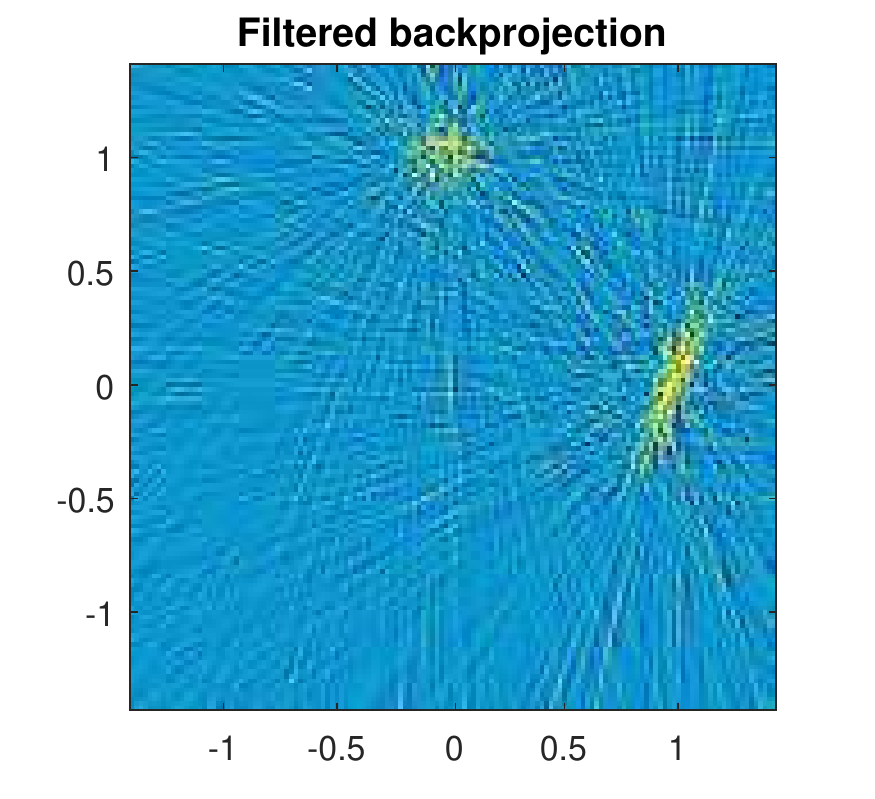}};
\end{minipage}%
    \hfill%
\begin{minipage}[t]{0.5\linewidth}
	{\includegraphics[width=\linewidth]{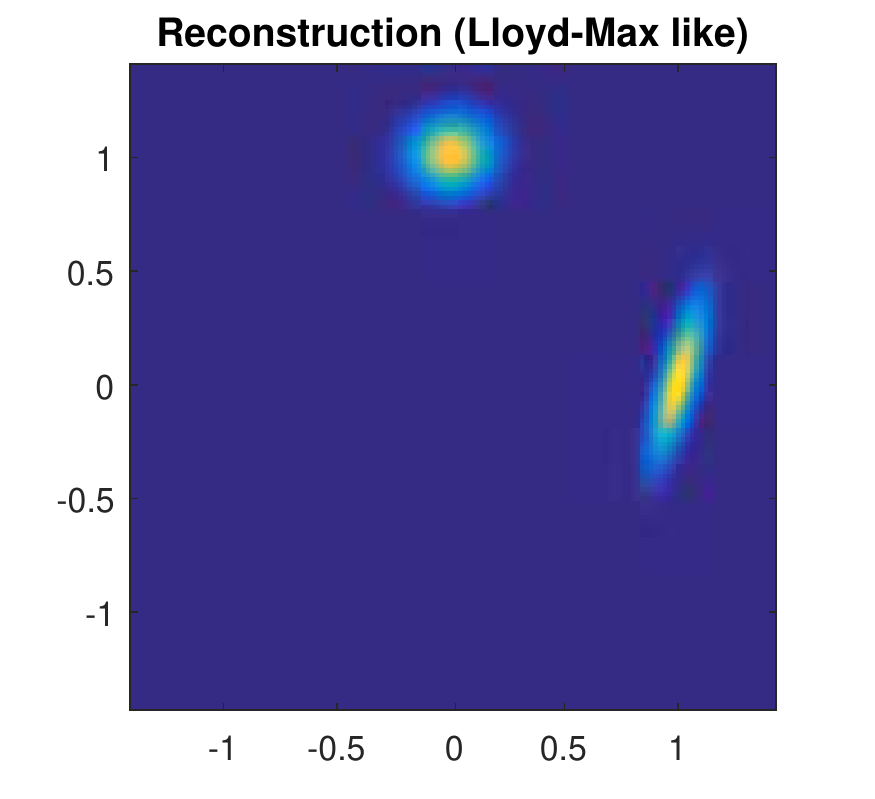}};
\end{minipage}
\caption{(\textbf{a}) Classical FBP reconstruction. (\textbf{b}) Proposed reconstruction using $L_1$ minimization.}
	\label{fig:reconstruction}
\end{figure}

For synthetic measurements from a variety of original (real) GMMs the algorithm proved robust regardless of the values of initial parameters, with estimation using $L_1$ minimization and the scaling factor $k$ correcting the bias. The $L_2$ minimization method proved inefficient, since wrongly assigned LORs would cause unstable estimations and "breaks" in the algorithm. An illustration of the results for $K=2$, $N=4000$ is shown in Fig.~\ref{fig:reconstruction}, along with the corresponding classical FBP method.

At the end, we would like to draw the readers attention to the fact that the reconstructed image is given by its parametric model: mean vectors $\{\boldsymbol{\mu}_k\}$, covariance matrices $\{\boldsymbol{\Sigma}_k\}$ and mixture weights $\{\tau_k\}$. It is virtually of infinite resolution, since the Gaussian components can be evaluated at each spatial point. The model is sparse: it consists from only a few parameters needed for the successful object representation. Hence, future research will be oriented to compressed sensing approach (\cite{rani2018systematic, Ralasic2018, Ralasic2019}) for reducing the number of projections, in this case reduced radiotracer’s concentration. Due to robustness of the proposed reconstruction method, post-filtering step is not needed. 

%

%

%



\bibliographystyle{IEEEtran}
\bibliography{myrefs}{}
\appendices{
\section{} \label{AppendixA}
We will find the solution to \eqref{eq:muproblem} using classical matrix calculus techniques. First note that the solution to \eqref{eq:xmin} is 
\begin{gather*}
\begin{bmatrix}
\bm{x}_i^m 	\\
\lambda	\\	
\end{bmatrix}
=
\begin{bmatrix}
2\bm{W} & \bm{a}_i  \\ 
\bm{a}_i^T & 0 
\end{bmatrix}^{-1}\cdot\begin{bmatrix}
2\bm{W}\bm{\mu}\\
-l_i
\end{bmatrix}.
\end{gather*}
To accommodate this, we will denote the expression in \eqref{eq:muproblem} by $d^2$ and expand it:
 \begin{gather*}\label{app:sum1}
 d^2=\sum_{i=1}^N(\bm{x}_i^m-\bm{ \mu})^T \bm{W}  (\bm{x}_i^m-\bm{\mu})\nonumber\\
 = \sum_{i=1}^N \left(\begin{bmatrix}
\bm{x}_i^m 	\\
\lambda	\\	
\end{bmatrix}-\begin{bmatrix}
\bm{\mu} 	\\
\lambda	\\	
\end{bmatrix}\right)^T\begin{bmatrix}
\bm{W} & \bm{0} 	\\
\bm{0}	& 0\\	
\end{bmatrix} \left(\begin{bmatrix}
\bm{x}_i^m 	\\
\lambda	\\	
\end{bmatrix}-\begin{bmatrix}
\bm{\mu} 	\\
\lambda	\\	
\end{bmatrix}\right),\nonumber\\
 \end{gather*}
 For simplicity, denote 
 \[ \bm{x}_{i\lambda}=\begin{bmatrix}
\bm{x}_i^m 	\\
\lambda	\\	
\end{bmatrix},\bm{\mu}_\lambda=\begin{bmatrix}
\bm{\mu} 	\\
\lambda	\\	
\end{bmatrix}\text{ and }\widetilde{\bm{W}}=\begin{bmatrix}
\bm{W} & \bm{0} 	\\
\bm{0}	& 0\\	
\end{bmatrix}.\]
Now $d^2$ equals
\begin{gather*}\label{app:sum2}
\sum_{i=1}^N \left(\bm{x}_{i\lambda}^T\widetilde{\bm{W}}\bm{x}_{i\lambda}-\bm{x}_{i\lambda}^T\widetilde{\bm{W}}\bm{\mu}_{\lambda}-\bm{\mu}_{\lambda}^T\widetilde{\bm{W}}\bm{x}_{i\lambda}+\bm{\mu}_{\lambda}^T\widetilde{\bm{W}}\bm{\mu}_{\lambda}\right),
\end{gather*}
which we will differentiate piecewise to find the minimum. We have:
\begin{itemize}[leftmargin=*]
\item[1)] \[\frac{d}{d\bm{\mu}}~\bm{x}_{i\lambda}^T\widetilde{\bm{W}}\bm{x}_{i\lambda}=2\begin{bmatrix}
2\bm{W}\bm{\mu} 	\\
-l_i	\\	
\end{bmatrix}^T \bm{B}_i\begin{bmatrix}
2W 	& \bm{a}_i  \\ 
\bm{a}_i^T & 0 
\end{bmatrix}^{-1}\begin{bmatrix}
2\bm{W} 	\\
\bm{0}	\\	
\end{bmatrix},\]
where
\[\bm{B}_i=\left(\begin{bmatrix}
2W 	& \bm{a}_i  \\ 
\bm{a}_i^T & 0 
\end{bmatrix}^{-1}\right)^T \widetilde{\bm{W}}.\]
\item[2)]\[\frac{d}{d\bm{\mu}}~\bm{\mu}_{\lambda}^T\widetilde{\bm{W}}\bm{x}_{i\lambda}=\bm{\mu}_\lambda^T \bm{B}_i^T\begin{bmatrix}
2\bm{W} 	\\
\bm{0}	\\	
\end{bmatrix}+\begin{bmatrix}
2\bm{W}\bm{\mu} 	\\
-l_i	\\	
\end{bmatrix}^T \bm{B}_i \begin{bmatrix}
\bm{I} 	\\
\bm{0}	\\	
\end{bmatrix}.\]
\item[3)]\[\frac{d}{d\bm{\mu}}~\bm{\mu}_{\lambda}^T\widetilde{\bm{W}}\bm{x}_{i\lambda}==\bm{\mu}_\lambda^T \bm{B}_i^T\begin{bmatrix}
2\bm{W} 	\\
\bm{0}	\\	
\end{bmatrix}+\begin{bmatrix}
2\bm{W}\bm{\mu} 	\\
-l_i	\\	
\end{bmatrix}^T \bm{B}_i \begin{bmatrix}
\bm{I} 	\\
\bm{0}	\\	
\end{bmatrix}\]
\item[4)]\[\frac{d}{d\bm{\mu}}~\bm{\mu}_{\lambda}^T\widetilde{\bm{W}}\bm{\mu}_{\lambda}=2\bm{\mu}_\lambda^T\widetilde{\bm{W}}\begin{bmatrix}
\bm{I} 	\\
\bm{0}	\\	
\end{bmatrix}.\]
Note that in all calculations we use the fact that $\bm{W}$ and, by extension, $\widetilde{\bm{W}}$ are symmetric. By plugging these equations into $\frac{d}{d\bm{\mu}}$ and equating that with $0$ to obtain the minimum, we get:
\begin{gather*}
\frac{d}{d\bm{\mu}}d^2=\sum_{i=1}^N \left(\begin{bmatrix}
2\bm{W}\bm{\mu} 	\\
-l_i	\\	
\end{bmatrix}^T \cdot \bm{B}_i\left(\begin{bmatrix}
2W 	& \bm{a}_i  \\ 
\bm{a}_i^T & 0 
\end{bmatrix}^{-1}\begin{bmatrix}
2\bm{W} 	\\
\bm{0}	\\	
\end{bmatrix}- \begin{bmatrix}
\bm{I} 	\\
\bm{0}	\\	
\end{bmatrix}\right)-\right.\\
\left.-\bm{\mu}_\lambda^T\left(\bm{B}_i^T\begin{bmatrix}
2\bm{W} 	\\
\bm{0}	\\	
\end{bmatrix}-\widetilde{\bm{W}}\begin{bmatrix}
\bm{I} 	\\
\bm{0}	\\	
\end{bmatrix}\right)\right)=0.
\end{gather*}
Define
\begin{gather*}
\begin{bmatrix}
\bm{M}_i 	\\
\bm{m}_i	\\	
\end{bmatrix}
=\bm{B}_i\left(
\begin{bmatrix}
2W 	& \bm{a}_i  \\ 
\bm{a}_i^T & 0 
\end{bmatrix}
\begin{bmatrix}
2\bm{W} 	\\
\bm{0}	\\	
\end{bmatrix}
- \begin{bmatrix}
1 & 0  	\\
0 & 1	\\	
0 & 0 \\
\end{bmatrix}\right),
\\
\begin{bmatrix}
\bm{N}_i 	\\
\bm{0}	\\	
\end{bmatrix}
=\bm{B}_i^T\begin{bmatrix}
2\bm{W} 	\\
\bm{0}	\\	
\end{bmatrix}-
\widetilde{\bm{W}}
\begin{bmatrix}
1 & 0  	\\
0 & 1	\\	
0 & 0 \\
\end{bmatrix}.
\end{gather*}
From the previous equation we now have 
\begin{gather*}
\sum_{i=1}^N \left([2\bm{\mu}^T\bm{W}^T, -l_i]\cdot \begin{bmatrix}
\bm{M}_i 	\\
\bm{m}_i	\\	
\end{bmatrix}-[\bm{\mu}^T, \lambda_i]\cdot \begin{bmatrix}
\bm{N}_i 	\\
\bm{0}	\\	
\end{bmatrix}
\right) = 0\\
\sum_{i=1}^N \left(2\bm{\mu}^T\bm{W}^T\bm{M}_i-l_i\bm{m}_i-\bm{\mu}^T\bm{N}_i\right)=0\\
\bm{\mu}^T\sum_{i=1}^N\left(2\bm{W}^T \bm{M}_i-\bm{N}_i\right)=\sum_{i=1}^Nl_i\bm{m}_i,
\end{gather*}
from which it follows that
\[\bm{\mu}^T=\left(\sum_{i=1}^Nl_i\bm{m}_i\right)\cdot \left(\sum_{i=1}^N\left(2\bm{W}^T \bm{M}_i-\bm{N}_i\right)\right)^{-1}.\]
It remains to verify that this stationary point $\bm{\mu}$ is also a turning point. However, since a point whose sum of squared distances from all lines is minimal must exist from a geometrical perspective, the solution in \eqref{eq:mu} is indeed the (global) minimum.
\end{itemize}
\section{} \label{AppendixB}
The $L_1$ minimization algorithm recursively reduces and increases dimensionality of the observed subspace and uses weighted median to efficiently find the global minimum. Reduction of dimensionality is achieved by extracting of parameters and inserting them into remaining equations in \eqref{eq:sigmaminL1}. 
If $[A_{i1}~ A_{i2}~ A_{i3}]$ is the $i$-th row of matrix $\bm{A}$, and $b_i$ the $i$-th element of vector $k\cdot \bm{b}$, the $i$-th equation of the system is 
\[b_i = \Sigma_{11} A_{i1} + \Sigma_{12} A_{i2} + \Sigma_{13} A_{i3}.\]
We choose equation $j_1$ from the set $i=1,...,N$ and extract one of its parameters, e.g. $\Sigma_{11}$:
\begin{equation}\label{eq:sigma_11}
\Sigma_{11} = -\frac{A_{j_1 2}}{A_{j_1 1}}\Sigma_{12} - \frac{A_{j_1 3}}{A_{j_1 1}}\Sigma_{22} + b_{j_1}.
\end{equation}

We insert it into all other equations and get a new system with only two unknown parameters:
\[b_i - b_{j_1}A_{i1} = \left(A_{i2} - \frac{A_{j_1 }}{A_{j_1 1}}A_{i1}\right) \Sigma_{12} + \left(A_{i3} -
\frac{A_{j_1 3}}{A_{j_1 1}}A_{i1}\right) \Sigma_{22},\]
for $ i\neq j_1.$
Now, we choose some other equation $j_2, j_2\neq j_1$ and extract one of the remaining parameters, e.g. $\Sigma_{12}$: 

\begin{equation}\label{eq:sigma_12}
\Sigma_{12} = -\frac{A_{j_2 3} A_{j_1 1} - A_{j_1 3} A_{j_2 1}}{A_{j_2 2} A_{j_1 1} - A_{j_1 2} A_{j_2 1}} \Sigma_{22}
+ \frac{\left(b_{j_2}-b_{j_1} A_{j_2 1}\right)A_{j_1 1}}{A_{j_2 2} A_{j_1 1} - A_{j_1 2} A_{j_2 1}}.
\end{equation}

We insert $\Sigma_{12}$ into all other equations and get the system of $N-2$ equations with only one unknown parameter $\Sigma_{22}$:
\begin{equation}\label{eq:new_bi}
b^{(1)}_i = A^{(1)}_i \Sigma_{22},
\end{equation}
where $i\neq j_1, j_2$,
\[b^{(1)}_i = b_i - b_{j_1}A_{i1} - \left(A_{i2} - \frac{A_{j_1 2}}{A_{j_1 1}}A_{i1}\right)
\frac{\left(b_{j_2}-b_{j_1} A_{j_2 1}\right)A_{j_1 1}} {A_{j_2 2} A_{j_1 1} - A_{j_1 2} A_{j_2 1}},\]

\begin{align*}
A^{(1)}_i &= A_{i3} - \frac{A_{j_1 3}}{A_{j_1 1}}A_{i1} - \\
&-\left(A_{i2} - \frac{A_{j_1 2}}{A_{j_1 1}}A_{i1}\right)
\frac{A_{j_2 3} A_{j_1 1} - A_{j_1 3} A_{j_2 1}}{A_{j_2 2} A_{j_1 1} - A_{j_1 2} A_{j_2 1}}.
\end{align*}

Parameter $\Sigma_{22}$ is calculated by minimizing $L_1$ norm

\begin{equation}\label{eq:min_L1}
\min \sum_{\substack{i=1\\i\neq j_1, j_2 }}^N 
\left|b^{(1)}_i- A^{(1)}_i\Sigma_{22} \right| =  \\
\min \sum_{\substack{i=1\\i\neq j_1, j_2 }}^N \left|A^{(1)}_i\right|
\left|\frac{b^{(1)}_i}{A^{(1)}_i} - \Sigma_{22} \right|.
\end{equation} 

The value of parameter $\Sigma_{22}$ is given by the weighted median (MED):
\begin{equation}\label{eq:sigma_22}
\left(\Sigma_{22}, j_3\right) = \text{MED} \left( \left|A^{(1)}_i\right|
\Diamond\frac{b^{(1)}_i}{A^{(1)}_i} \biggm|_{i=1, i\neq j_1, j_2}^N
\right),
\end{equation}
where $j_3$ is an ordinal number of the concomitant equation and $\Diamond$ is the replication operator. The weighted median can be obtained using the algorithm given in \cite{Sovic2018}. The value of parameter $\Sigma_{22}$ is an element of set $\{b^{(1)}_i / A^{(1)}_i \}$. Chosen equations $\{j_1, j_2, j_3\}$ define a local minimum in 1D, a vertex. 

Return to a higher dimension is achieved by putting calculated parameter value $\Sigma_{22}$ in \eqref{eq:sigma_12}. 
To further descending in $L_1$ cost surface, we fix equations $j_1$ and $j_3$, and try to find new $j_4$ using \eqref{eq:new_bi}-\eqref{eq:sigma_22}. If $j_4\neq j_2$, new vertex is defined by $\{j_1, j_2, j_3\} = \{j_1, j_3, j_4\}$ and we repeat the same procedure. If $j_4=j_2$, we conclude that the vertex $\{j_1, j_2, j_3\}$ is a local minimum in observed 2D subspace, thus we return to 3D by fixing $j_2$ and $j_3$ and try to find new $j_4$ instead of $j_1$. We repeat the previous procedure until we cannot find new equation. Since the $L_1$ cost surface is convex, the global minimum is reached. Parameters $\Sigma_{11}$, $\Sigma_{12}$ and $\Sigma_{22}$ are given by \eqref{eq:sigma_11}, \eqref{eq:sigma_12} and \eqref{eq:sigma_22}.
}
\end{document}